\documentclass[12pt]{iopart}
\usepackage{graphicx}
\usepackage[english]{babel}

\begin{document}

\title{Cosmological models with Gurzadyan-Xue dark energy}
\author{G. V. Vereshchagin\footnote{%
e-mail address: \texttt{veresh@icra.it}} and G. Yegorian\footnote{%
e-mail address: \texttt{gegham@icra.it}}}
\address{ICRANet, P.le della Repubblica 10, I65100 Pescara, Italy and ICRA, Dip.
Fisica, Univ. \textquotedblleft La Sapienza\textquotedblright , P.le A. Moro
5, I00185 Rome, Italy}

\begin{abstract}
The formula for dark energy density derived by Gurzadyan and Xue is the only formula
which provides (without a free parameter) a value for dark energy density 
in remarkable agreement with current cosmological datasets, unlike numerous phenomenological
scenarios where the corresponding value is postulated. This formula suggests
the possibility of variation of physical constants such as the speed of light
and the gravitational constant. Considering several cosmological models
based on that formula and deriving the cosmological equations for each case, we show
that, in all models source terms appear in the continuity
equation. So, on the one hand, GX models make up a rich set covering a lot of 
currently proposed models of dark energy, on the other hand, they reveal 
hidden symmetries, with a particular role of the separatrix $\Omega_m=2/3$, and 
link with the issue of the content of physical constants.
\end{abstract}



\maketitle

\section{Introduction}

With the establishment of observational evidence favoring dark energy
domination in the recent history of the cosmological expansion, numerous
different models were suggested to account for it. For most of them the
value of the corresponding density parameter is chosen to fit the
observational one. In contrast, Gurzadyan and Xue formula \cite{GX} for dark
energy predicts the observed value for the density parameter of the dark
energy. The formula reads \cite{GX,GX2}
\begin{equation}
\rho _{GX}=\frac{\pi }{8}\,\frac{\hbar c}{L_{p}^{2}}\,\frac{1}{a^{2}}=\frac{%
\pi }{8}\,\frac{c^{4}}{G}\,\frac{1}{a^{2}},  \label{rhoLambda}
\end{equation}%
where $\hbar $ is the Planck constant, the Planck length is $L_{p}=\left( 
\frac{\hbar G}{c^{3}}\right) ^{\frac{1}{2}}$, $c$ is the speed of light, $G$
is the gravitational constant, $a$ is the scale factor of the Universe.
Following the original idea by Zeldovich \cite{zelvac}, it corresponds to
the cosmological term 
\begin{equation}
\Lambda _{GX}=\frac{8\pi G\rho _{GX}}{c^{2}}=\frac{8\pi G}{c^{2}}\frac{\pi }{%
8}\,\frac{c^{4}}{G}\,\frac{1}{a^{2}}=\pi ^{2}\left( \frac{c}{a}\right) ^{2}.
\label{Lambda}
\end{equation}%
Therefore, to keep it constant, the speed of light should vary with
cosmological expansion as $c\propto a$. One can consider other possibilities
as well, assuming the presence of some fundamental physical quantity and
admitting variation of basic physical constants \cite{Ver05} in the spirit
of Dirac approach. Since the Planck constant does not appear in (\ref%
{rhoLambda}), these can be the gravitational constant and the speed of light.

In the literature there is a long-standing discussion how to understand
possible variability of physical constants, linking this point to the
meaning of world constants and units in physics \cite{Duf02}. Clearly
experimentally only dimensionless constants variation can be detected, but
the meaning and the content of the underlying physical theory changes
depending on the issue which dimensionful quantity varies.

In fact, models with varying physical constants are among the currently
discussed ones (e.g. \cite{B}). However, models based on the formula (\ref%
{rhoLambda}) are very different from various phenomenological models of dark
energy, phantoms etc., because of the underlying \emph{empirical basis},
namely agreement between predictions for dark energy density parameter and
the value, following from the array of recent observations.

The classification of various cosmological models following from GX-formula
is given in \cite{Ver05}. Furthermore, it is shown \cite{VY06b} that GX
models provide indeed good fit to supernovae and radio galaxies data. These
models contain hidden invariance, the separatrix dividing cosmological
solutions into two classes: singular Friedmannian-type and non-singular with
vanishing initial density \cite{VY06b},\cite{VY06a}. In this paper we
provide cosmological equations for the simplest cases discussed in \cite%
{Ver05} and generalize them, including radiation field, and compare them to
the standard cosmological model.

In the section 2 cosmological framework with varying physical constants is
formulated. In the section 3 we derive cosmological equations for particular
cases, discussed in \cite{Ver05}. We provide analytical solutions for energy
density in all models, and also in some models for the scale factor.
Conclusions follow in the last section.

\section{Cosmological equations with varying constants}

In this section we are going to derive cosmological equations, assuming
varying constants, such as the speed of light $c(t)$, the gravitational
constant $G(t)$ and the cosmological term $\Lambda(t)$ depending on the
cosmic time $t$. Einstein equations are postulated to be valid in such a
case.

We start from the FRW interval 
\begin{equation}
ds^{2}=-c(t)^{2}dt^{2}+a^{2}(t)\left[ \frac{dr^{2}}{1-kr^{2}}+r^{2}\left(
d\theta ^{2}+\sin ^{2}\theta d\varphi ^{2}\right) \right] ,  \label{int}
\end{equation}%
where $r,\theta ,\varphi $ are spatial coordinates, $%
a$ is the scale factor. Using Einstein equations 
\[
R_{\mu \nu }-\frac{1}{2}Rg_{\mu \nu }-\frac{\Lambda }{c^{2}}g_{\mu \nu }=%
\frac{8\pi G}{c^{4}}T_{\mu \nu }, 
\]%
and the energy-momentum tensor 
\[
T^{\mu }{}_{\nu }=diag(-\epsilon ,p,p,p), 
\]%
with $\rho $ being the energy density, $p$ being the pressure, we arrive at
the cosmological equations 
\begin{equation}
H^{2}+\frac{kc^{2}}{a^{2}}-\frac{\Lambda }{3}=\frac{8\pi G}{3c^{2}}\rho ,
\label{fe1}
\end{equation}%
\begin{equation}
\dot{H}+H^{2}-\frac{\Lambda }{3}=-\frac{4\pi G}{3c^{2}}\left( \rho
+3p\right) +H\frac{\dot{c}}{c},  \label{fe2}
\end{equation}%
where $k$ is the sign of spatial curvature, $H\equiv d\ln a/dt$ is the
Hubble parameter, a dot denotes time derivative. From (\ref{fe1}) and (\ref%
{fe2}) (or from $\frac{8\pi G}{c^4}T^{\mu \nu }{}_{;\nu }=0$) continuity equation can be
written 
\begin{equation}
\dot{\rho}+3H\left( \rho +p\right) =-\dot{\rho}_{\Lambda }+\left( \rho +\rho
_{\Lambda }\right) \left( 4\frac{\dot{c}}{c}-\frac{\dot{G}}{G}\right) ,
\label{mue}
\end{equation}

where have we introduced $\rho _{\Lambda }=\Lambda c^{2}/(8\pi G)$. The same
equations were derived e.g. in \cite{Har99}. The difference with \cite{Har99}
is that we consider the scaling (\ref{rhoLambda}) not as a phenomenological,
but as an empirical relation.

It is easy to check that (\ref{fe1}) is the first integral of (\ref{fe2}) as
it is within standard Friedmann cosmology. Cosmological equations for the
case when $c$ and $G$ both depend on time can be found in \cite{Alb99}. For
the case when instead $\Lambda $ and $G$ depend on time these equations are
given in \cite{Sha05}. Our equations reduce to equations of \cite{Alb99} and 
\cite{Sha05} with the corresponding assumptions. There is a little
difference with \cite{Alb99}, because we postulated only invariance of the
Einstein equations, while authors of \cite{Alb99} assumed that both Einstein
and Friedmann equations are still valid. With these assumptions the last
term in (\ref{fe2}) disappears and our equations looks identical to those of 
\cite{Alb99}.

Note, that the former energy conservation does not hold any more: there are
sources and drains of energy (e.g. from dark energy to the usual matter,
from the variation of the speed of light and the gravitational field
strength) in (\ref{mue}). As we will see in the next section, (\ref{mue})
can be integrated to obtain energy density as explicit function of the scale
factor. Further for some models Friedmann equation (\ref{fe1}) can be
explicitly integrated to get the scale factor as a function of time.

In \cite{Ver05} five different cases have been discussed, namely:

\begin{enumerate}
\item {c=const, G=const, $\Lambda\neq$const;}

\item {$c\propto a$, G=const, $\Lambda=$const;}

\item {c=const, $G\propto a^{-2}$, }${\rho }${$_{\Lambda }/c^{2}=$const, $\Lambda\neq$const
(case 3.1);}

\item {$c\propto a^{1/2}$, G=const, }${\rho }${$_{\Lambda }=$const, $\Lambda\neq$const (case
4.2);}

\item {$c\neq $const, $G\neq $const, $\Lambda \neq $const.}
\end{enumerate}

In what follows we analyze each of the first four cases, provide cosmological equations and corresponding solutions for them.

\section{Cosmological models based on Gurzadyan-Xue formula}

In this section we present cosmological equations for each of the cases
(i-iv) and discuss the corresponding cosmological scenarios, providing
analytical solutions, where possible.

\subsection{Models with dark energy and pressureless matter}

In this subsection we put the pressure in (\ref{fe2}) and (\ref{mue}) to
zero and introduce mass density $\mu =\rho /c^{2}$. Each solution can be
parametrized by the usual matter density

\begin{eqnarray}
\Omega _{m}\equiv\frac{\mu _{0}}{\mu _{c}}\equiv\frac{8\pi G_{0}}{3H_{0}^{2}}\mu _{0}, 
\label{muc}
\end{eqnarray}
where $\mu_c$ is the critical density in usual Friedmannian models, subscript 0 denotes the value of each quantity today. In what follows, to define density parameters we use normalization condition

\begin{eqnarray}
\sum_{i}\Omega_i=1,
\label{norm}	
\end{eqnarray}
where $i$ numbers components such as dark energy, radiation, usual matter.

In order to get density as a function of the scale factor we rewrite (\ref%
{mue}) as

\begin{equation}
\frac{d\mu }{da}+3\frac{\mu }{a}=\textrm{R}\equiv\frac{1}{4\pi G}\,\frac{dc}{cda}\left( 8\pi G\mu +\Lambda\right)-\mu \frac{dG}{Gda}-\frac{1}{8\pi G}\,\frac{d\Lambda }{da}. 
\label{cont}
\end{equation}

The crucial feature of these models is the presence of a separatrix in
solutions for the density as a function of the scale factor\ (see \cite%
{VY06b} and \cite{VY06a}) . This separatrix divides solutions in two
classes: the one with Friedmannian singularity, and the one with vanishing
density and non-zero scale factor in the beginning.

To see the difference from our models and the corresponding Friedmannian ones we compute for each case \emph{mass transfer rate}, namely the ratio between R of (\ref{cont}) and expansion term $3\mu/a$, i.e.
\begin{equation}
{\mathcal M}\equiv\frac{a R}{3\mu}.
\label{mtr}
\end{equation}

If dimensionless ${\mathcal M}$ vanishes, it means the usual Friedmannian solution $\mu\propto a^{-3}$ holds. When ${\mathcal M}=1$ mass transfer cancels the effect of expansion and the mass density does not change in course of expansion. If it is large, effect of expansion can be neglected.

\textbf{Model 1 (varying cosmological constant).} Neither the speed of light nor the gravitational constant
vary with time, but in this case from (\ref{Lambda}) we find $\Lambda
\propto a^{-2}$. Cosmological equations are

\[
H^{2}+\frac{k^{\prime }c^{2}}{a^{2}}=\frac{8\pi G}{3}\mu , \nonumber
\]
\[
\dot{\mu}+3H\mu =\frac{\pi }{4G}\left( \frac{c}{a}\right) ^{2}H, 
\]%
where $k^{\prime }=k-\frac{\pi ^{2}}{3}$. It is clear that $k^{\prime }<0$,
i.e. the effect of $\Lambda $ variation is to make spatial geometry
negatively curved, even if the original sign of $k$ is $"+"$. The source
term in the continuity equation can be interpreted as the energy transfer
from the dark energy component into the matter component (see \cite{Sha05}
and references therein).

Solution for the density is given by

\begin{equation}
\mu (a)=\mu _{0}\left( \frac{a_{0}}{a}\right) ^{3}+\frac{\pi }{4G}\,\frac{%
c^{2}}{a^{2}}\left( 1-\frac{a_{0}}{a}\right) ,  \label{mumod1}
\end{equation}
and after substitution of this solution to (\ref{fe1}) we obtain

\begin{equation}
	\begin{array}{l}
	\pm A_{1}^\frac{3}{2}t=\sqrt{a A_1(a A_1+B_1)}-\sqrt{a_0 A_1(a_0 A_1+B_1)}+ \\ \displaystyle{\quad\quad\quad\quad\quad\quad\quad\quad\quad\quad\quad\quad\quad +B_1\log\left(\frac{\sqrt{a_0 A_1}+\sqrt{a_0 A_1+B_1}}{\sqrt{a A_1}+\sqrt{a A_1+B_1}}\right)},
	\label{sol1}
	\end{array}
\end{equation}
where 
\[
A_{1}=c^{2}(\pi ^{2}-k),\quad B_{1}=\frac{2\pi }{3}a_{0}(4G\mu
_{0}a_{0}^{2}-\pi c^{2}). 
\]%
This solution reduces to a simple form for separatrix value with the
condition $B_{1}=0$, namely%
\begin{equation}
a=a_{0}+\sqrt{A_{1}}t.  \label{ballistic}
\end{equation}

The separatrix solution is given by the density parameter $\Omega
_{m}\approx 2/3$ and depends slightly on the spatial curvature. For high
values of the density parameter solutions practically coincide with the
corresponding Friedmannian ones. In such a case one can neglect the source
term in the continuity equation. This also can be done for large scale
factors since solutions approach the separatrix where they describe
Friedmann Universe with negative curvature.

The present scale factor is well defined quantity even for $k=0$ models,
\begin{equation}
a_0=(-k')^\frac{1}{2}\frac{c}{H_0}\frac{1}{\sqrt{1-\Omega_m}},
\label{a01}
\end{equation}
since this is the model with effective negative curvature with $\Omega_k=-\frac{k'c^2}{a_0^2 H_0^2}$. With this definition (\ref{norm}) is satisfied along with with (\ref{muc}).

Mass transfer rate (\ref{mtr}) for this model in terms of density parameter is given by
\[
{\mathcal M}=\frac{2\pi^2}{9k'}\left(\frac{1}{\Omega_m}-1\right),
\]
and it is clear that ${\mathcal M}$ gets infinite with $\Omega_m\rightarrow 0$, i.e. vacuum mass transfer dominates expansion, and in the opposite case the model approaches Friedmannian matter dominated one with negative curvature.

\textbf{Model 2 (varying speed of light).} To keep interpretation of $\Lambda $ as a cosmological
constant, we have to require the varying speed of light, $c=\left( \frac{%
\Lambda }{\pi ^{2}}\right) ^{1/2}a$. At the same time the gravitational
constant does not change, $G$=const. Cosmological equations reduce to the
following system

\[
H^{2}-\frac{\Lambda ^{\prime }}{3}=\frac{8\pi G}{3}\mu , 
\]%
\[
\dot{\mu}+3H\mu =\frac{3H}{4\pi G}\left( H^{2}+\frac{k\Lambda }{\pi ^{2}}%
\right) =3H\mu \left( 1+\frac{\Lambda }{8\pi G\mu}\right) , 
\]%
where $\Lambda ^{\prime }=\Lambda \left( 1-\frac{3k}{\pi ^{2}}\right) $.
The source in the continuity equation now comes from the speed of light
variability.

There is complete analytical solution for this model 
\begin{equation}
\mu (a)=\mu _{0}\frac{a_{0}}{a}+\frac{\Lambda }{4\pi G}\left( 1-\frac{a_{0}}{%
a}\right) ,  \label{musol2}
\end{equation}%
\[
{a(t)=-\frac{A_{2}}{2B_{2}}+\frac{A_2+2a_{0}B_{2}}{2B_{2}}\cosh \left( t\sqrt{B%
}_{2}\right) \pm \sqrt{\frac{a_{0}}{B_{2}}\left( A_{2}+a_{0}B_{2}\right) }%
\sinh \left( t\sqrt{B}_{2}\right) }, 
\]%
where

\[
A_{2}=\frac{2}{3}a_{0}\left( 4\pi G\mu _{0}-\Lambda \right) ,\quad
B_{2}=\Lambda \left( 1-\frac{k}{\pi ^{2}}\right) . 
\]

In this model the separatrix is given by $\Omega _{m}=2/3$ independent on
the curvature. If one neglects the source in the continuity equation (that is
again possible for large $a$ or large $\Omega _{m}$), which consists of two
terms, proportional to $H^{3}$ and $H\Lambda $ respectively, we come to the
usual Friedmann-Lema\^{\i}tre cosmology with the cosmological constant\footnote{Note that the cosmological constant appears naturally within McCrea-Milne approach \cite{Gur85}.} $%
\Lambda ^{\prime }$.

The density parameter for $\Lambda$ term is
\[
\Omega_\Lambda=\frac{\Lambda}{3H_0^2}\left(1-\frac{3k}{\pi^2}\right),
\]
and the scale factor today is arbitrary quantity in this model.

Mass transfer rate (\ref{mtr}) is simply
\[
{\mathcal M}=\frac{1}{\Omega_m},
\]
so it precisely cancels expansion with $\Omega_m\rightarrow 1$.

\textbf{Model 3 (varying gravitational constant).} We introduce the fundamental constant $\mu_{GX}=\frac{\Lambda}{8\pi G}$=const, the density of the vacuum. In this case the
gravitational constant changes, $G=\frac{\pi}{8\mu_{GX}}\left(\frac{c}{a}%
\right)^2$, while the speed of light does not. The cosmological term also
acquires dependence on the scale factor through $G$ and is given by (\ref%
{Lambda}). Cosmological equations are

\begin{equation}
H^{2}+\frac{kc^{2}}{a^{2}}=\frac{\pi ^{2}}{3}\left( \frac{c}{a}\right)
^{2}\left( 1+\frac{\mu }{\mu _{GX}}\right) ,  \nonumber
\end{equation}%
\[
\dot{\mu}+3H\mu =2H\mu \left( 1+\frac{\mu _{GX}}{\mu }\right) , 
\]

These are very different from the usual Friedmann equations. Even if could
neglect the source terms in the continuity equation (that obviously cannot be done), the first cosmological
equation does not correspond to the first Friedmann one. However, phase
portrait of solutions for this model on the plane $\{\mu ,a\}$ looks
similar to the one of model 2 \cite{VY06a}.

Again there is solution for density 
\[
\mu =\mu _{0}\frac{a_{0}}{a}+2\mu _{GX}\left( 1-\frac{a_{0}}{a}\right) . 
\]%
Notice similarity with (\ref{musol2}). This is due to the fact that $\Lambda $ is a constant in the model 2, while $\mu _{GX}$ is constant in the model 3, so the dependence of matter density on the scale factor is the same, although the dependence of dark energy on the scale factor is different.

Solution for the scale factor looks instead exactly the same as the solution
for model 1, (\ref{sol1}) with 
\[
A_{3}=A_{1},\quad B_{3}=\frac{\pi ^{2}}{3\mu _{GX}}c^{2}a_{0}(\mu _{0}-2\mu
_{GX}). 
\]

The separatrix is the same as in model 1, given by $B_{3}=0$.

Matter density parameter for this model is
\[
\Omega_m=\frac{\pi^2}{3}\,\frac{c^2}{H_0^2 a_0^2}\,\frac{\mu}{\mu_{GX}},
\]
that is the same as (\ref{muc}). The scale factor today is the same as in model 1 and is given by (\ref{a01}), as this expression follows from the definition of density parameter for curvature, which is the same as in model 1. Another, equivalent expression follows from definitions of vacuum mass density and critical density and it gives
\[
a_0^2=\frac{\pi c^2}{8G_0\mu_{GX}}.
\]

Mass transfer rate (\ref{mtr}) for this model is
\[
{\mathcal M}=\frac{2}{3\Omega_m},
\]
that is analogous to the model 2, but with $\Omega_m\rightarrow 1$ it corresponds to a constant fraction of expansion rate.

\textbf{Model 4 (varying speed of light, constant vacuum energy density).} The constant vacuum energy density is introduced as $\rho
_{GX}=\frac{\Lambda c^{2}}{8\pi G}$=const, requiring variation of the speed
of light $c=\left( \frac{8G\rho _{GX}}{\pi }\right) ^{1/4}a^{1/2}$. Here $G$%
=const and $\Lambda =2\sqrt{2}\pi ^{3/2}\sqrt{G\rho _{GX}}a^{-1}=\frac{3\beta }{%
a\left( 1-\frac{3k}{\pi ^{2}}\right) }$, where the constant $\beta $
appears in the cosmological equations

\begin{equation}
H^{2}=\frac{8\pi G}{3}\mu +\frac{\beta }{a},  \nonumber
\end{equation}%
\[
\dot{\mu}+3H\mu =\frac{3H}{8\pi G}\left( H^{2}+\frac{\beta }{a}\,\frac{\pi
^{2}+3k}{\pi ^{2}-3k}\right) , 
\]%
with $\beta =\frac{2\sqrt{2}\pi ^{3/2}}{3}\left( G\rho _{GX}\right) ^{1/2}\left( 1-%
\frac{3k}{\pi ^{2}}\right) $.

The source term in the continuity equation can be neglected for large $a$,
but the first Friedmann equation contains the positive $a^{-1}$ term which
is again very different from the standard cosmology.

There is complete solution for this model with 
\[
\mu =\mu _{0}\left( \frac{a_{0}}{a}\right) ^{2}+\frac{1}{a}\sqrt{\frac{\pi
\rho _{GX}}{2G}}\left( 1-\frac{a_{0}}{a}\right) , 
\]%
and 
\[
a=a_{0}\pm t\sqrt{A_{4}+a_{0}B_{4}}+\frac{B_{4}}{4}t^{2}, 
\]%
where 
\[
A_{4}=\frac{8\pi G}{3}a_{0}\left( \mu _{0}a_{0}-\sqrt{\frac{\pi \rho _{GX}}{2G%
}}\right) ,\quad B_{4}=2\sqrt{\frac{2G\rho _{GX}}{\pi }}(\pi ^{2}-k). 
\]

This solution reduces to (\ref{ballistic}) for $B_{4}=0$ or, in other words,
when $\Omega _{m}\rightarrow 1$. The separatrix is given by $\Omega _{m}=2/3$%
, like in model 2.

Here new density parameter for dark energy is defined as
\[
\Omega_\beta=\frac{\beta}{H_0^2 a_0},
\]
and the scale factor value today follows from this definition
\[
a_0=\frac{c_0}{H_0}\frac{\pi}{\sqrt{3(1-\Omega_m})}.
\]
Notice that it is different from the scale factor in models 1 and 3 by the factor $\sqrt{1-3k/\pi^2}$.

Mass transfer rate (\ref{mtr}) for this model is given by
\[
{\mathcal M}=\frac{1}{3\Omega_m}+\left(1+\frac{3k}{\pi^2}\right)\left(\frac{1}{\Omega_m}-1\right),
\]
from which we find that with $\Omega_m\rightarrow 0$ the vaccum again dominates expansion, while in the opposite case this is just a constant fraction of expansion rate. Therefore, it cannot be reduced to the usual Friedmannian model.

Finally, in the case 5 of \cite{Ver05} cosmological equations are given by the system
(\ref{fe1})-(\ref{mue}). Definitions (\ref{muc}) and (\ref{norm}) can be used in this general case as well.

\subsection{Models with dark energy, dust and radiation}

So far we considered cosmological models based in GX formula (\ref{rhoLambda}%
), containing only dust and dark energy. In this subsection we turn to a
more general class of models, including into consideration radiation.

It is known (e.g. \cite{Har99}) that there is ambiguity in definition of
multicomponent models with dark energy and/or variation of constants. There
exist strong bounds on coupling between dark energy and radiation \cite%
{Oph04},\cite{Oph05}, so we avoid this ambiguity assuming decoupled
radiation, namely

\begin{equation}
\rho _{r}=\rho _{r0}\left( \frac{a}{a_{0}}\right) ^{-4},\quad p_{r}=\frac{%
\rho _{r}}{3},  \label{eqstate}
\end{equation}%
\bigskip where the subscript $"r"$ refers to radiation. In standard
cosmology any relativistic particle in thermodynamic equilibrium can be
described by these equations. However, varying speed of light implies
variation of dispersion relation $\epsilon ^{2}=\Pi c^{2}+m_{0}c^{4}$, where 
$\epsilon $ and $\Pi $ are particle's energy and momentum and $m_{0}$ is its
rest mass, hence influences equation of state for ultrarelativistic fluid.
In this paper we constrain ourselves with the case of thermal photons only.

Radiation density parameter in what follows is defined by
\begin{equation}
\Omega_r=\frac{8\pi G_0}{3H_0^2 c_0^2}\rho_{r0}.
\label{rhor}
\end{equation}

Then total energy conservation (\ref{mue}) implies

\begin{equation}
	\begin{array}{l}
	\displaystyle{\frac{d\mu }{da}+3\frac{\mu }{a}=\frac{\Lambda }{8\pi G}\left( 2\frac{dc}{cda%
}-\frac{d\Lambda }{\Lambda da}\right) +} \\ \displaystyle{\quad\quad\quad\quad\quad\quad\quad\quad\quad +\left[ \mu +\frac{\rho _{r0}}{c^{2}}%
\left( \frac{a}{a_{0}}\right) ^{-4}\right] \left( 4\frac{dc}{cda}-\frac{dG}{%
Gda}\right) -2\mu \frac{dc}{cda}.}
	\label{mue2}
	\end{array}
\end{equation}

It is clear from (\ref{mue2}) that in general matter couples to radiation
always except for the case $c=G$=const.

In this section we provide solutions for the set of cosmological models
described by (\ref{fe1}),(\ref{fe2}),(\ref{eqstate}) and (\ref{mue2}). Generally speaking, these
models do not reduce to the ones considered above because of special role of
radiation, given by (\ref{eqstate}).

\textbf{Model 1, with radiation.} This is the unique case when the radiation
does not couple to matter density, because neither speed of light, nor the
gravitational constant vary.

Cosmological equations for this model read

\begin{equation}
H^{2}+\frac{k^{\prime }c^{2}}{a^{2}}=\frac{8\pi G}{3}\left( \mu +\frac{\rho
_{r0}}{c^{2}}\left( \frac{a}{a_{0}}\right) ^{-4}\right) ,  \nonumber
\end{equation}%
\[
\dot{\mu}+3H\mu =\frac{\pi }{4G}\left( \frac{c}{a}\right) ^{2}H, 
\]

Solution for the matter density is the same (\ref{mumod1}) as it is in the
absence of radiation.

Additional peculiarity of this case is the presence of separatrix, playing
the same role as for models without radiation. The separatrix is defined by

\[
\Omega _{m}=\frac{2}{3}\frac{1-\Omega _{r}}{1-\frac{k}{\pi ^{2}}}, 
\]

that again gives $\Omega _{m}\approx 2/3$ for any curvature parameter.

Implicit solution for the scale factor as a function of cosmic time is

\[
t-t_{0}=\frac{1}{\tilde{A}_{2}}\left( \sqrt{F(a)}-\sqrt{F(a_{0})}\right) -%
\frac{\tilde{A}_{1}}{2\tilde{A}_{2}^{\frac{3}{2}}}\log \left( \frac{\tilde{A}%
_{1}+2a\tilde{A}_{2}+2\sqrt{\tilde{A}_{1}F(a)}}{\tilde{A}_{1}+2a_{0}\tilde{A}%
_{2}+2\sqrt{\tilde{A}_{1}F(a_{0})}}\right) , 
\]

where

\[
F(x)=\tilde{A}_{0}+\tilde{A}_{1}x+\tilde{A}_{2}x^{2}, 
\]

and%
\begin{eqnarray*}
\tilde{A}_{0} &=&\frac{8\pi G}{3c^{2}}\rho _{r0}a_{0}^{4}, \\
\tilde{A}_{1} &=&\frac{8\pi G}{3}a_{0}\left( \mu _{0}a_{0}^{2}-\frac{\pi
c^{2}}{4G}\right) , \\
\tilde{A}_{2} &=&\left( \pi ^{2}-k\right) c^{2}.
\end{eqnarray*}

The solution for separatrix is explicit and looks rather simple

\[
a=\frac{c}{H_{0}}\sqrt{\frac{\pi ^{2}-k}{1-\Omega _{r}}}\sqrt{%
1+2H_{0}t+(1-\Omega _{r})H_{0}^{2}t^{2}}. 
\]
\bigskip

\begin{figure}[htp]
	\centering
		\includegraphics{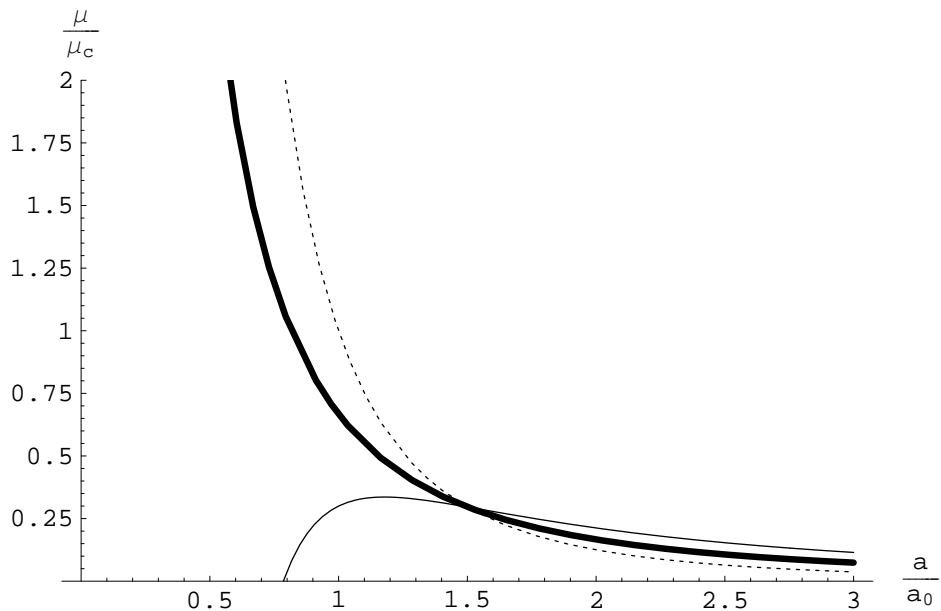}
\caption{Density of matter in model 1 in
terms of critical density, depending on the scale factor in terms of present
scale factor with $\Omega _{m}$ equal to 0.3 (firm curve), 0.667 (thick
curve) and 0.999 (dotted curve).}
\label{mod1}
\end{figure}

This solution approaches (\ref{ballistic}) at late times. Solution for
density is represented at fig. (\ref{mod1}) for selected values of $\Omega
_{m}$. Like in models without radiation solutions with $\Omega _{m}>2/3$
reach Friedmannian singularity in the past, while solutions with $\Omega
_{m}<2/3$ start with vanishing density.

Both density parameter and critical density are defined by (\ref{muc}). Radiation density parameter is as usual (\ref{rhor}). Curvature density parameters remains the same as in model without radiation, and the scale factor becomes a little bigger due to the change $\sqrt{1-\Omega_m}\rightarrow\sqrt{1-\Omega_m-\Omega_r}$ in (\ref{a01}).

\textbf{Model 2, with radiation.} Since the speed of light varies in this
model, matter becomes coupled to radiation and solution gets complicated.

Cosmological equations for this model are

\begin{equation}
H^{2}-\frac{\Lambda ^{\prime }}{3}=\frac{8\pi G}{3}\left( \mu +\frac{\pi^2}{\Lambda}\rho
_{r0}\,\frac{a_0^4}{a^6}\right) ,  \nonumber
\end{equation}%
\[
\dot{\mu}+H\left( \mu -\frac{\Lambda }{4\pi G}\right) =\frac{4\pi ^{2}\rho
_{r0}a_{0}^{4}}{\Lambda }\frac{H}{a^{6}}. 
\]

Hence the solution for the density is

\[
\mu (a)=\frac{1}{a}\left[ \frac{\Lambda }{4\pi G}\left( a-a_{0}\right) +\mu
_{0}a_{0}+\frac{4\pi ^{2}}{5\Lambda a_{0}}\rho _{r0}\left( 1-\left( \frac{a}{%
a_{0}}\right) ^{-5}\right) \right] . 
\]

\begin{figure}[htp]
	\centering
		\includegraphics{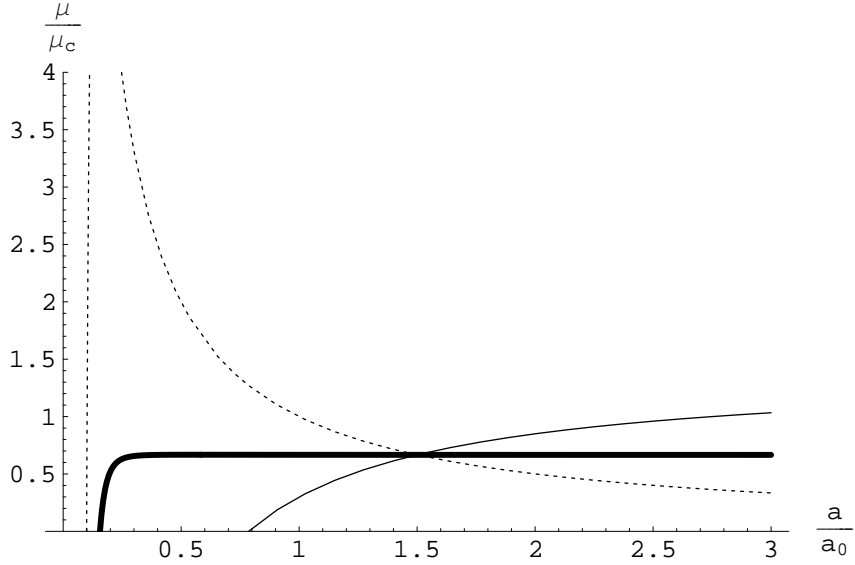}
\caption{Density of matter in model 2. $\Omega _{m}$ equals to 0.3 (firm curve), 0.667 (thick
curve) and 0.999 (dotted curve).}
\label{mod2}
\end{figure}

This model does not possess separatrix in the sense discussed above, because
the early expansion is dominated by the negative definite term (since the
vacuum energy is positive definite). The solution for matter density in this model is represented at fig. (\ref{mod2}).

Although there is no separatrix with respect to initial singularity, the
value $\Omega _{m}=2/3$ again plays important role, dividing solutions in
two classes, depending on the sign of time derivative of the density: for $%
\Omega _{m}>2/3$ the density decreases with time in matter dominated epoch,
like in Friedmannian models, while for $\Omega _{m}<2/3$ the density
increases with time. This behavior is seen from equation for the density,
since the condition $\Lambda =4\pi G\mu $ leads to $\Omega _{m}=2/3$.

In contrast to simple GX model without radiation, the scale factor is no more arbitrary in this model and is given by (\ref{a01}) with the change $\sqrt{1-\Omega_m}\rightarrow\sqrt{1-\Omega_m-\Omega_r}$. Radiation density parameter is
\[
\Omega_r=\frac{\pi^2}{\Lambda}\,\frac{\rho_{r0}}{\mu_c a_0^2},
\]
and it coincides with (\ref{rhor}).

\textbf{Model 3, with radiation.} Cosmological equations now reduce to

\begin{equation}
H^{2}+\frac{kc^{2}}{a^{2}}=\frac{2\pi ^{2}}{3}\left( \frac{c}{a}\right) ^{2}%
\left[ 1+\frac{1}{\mu _{GX}}\left( \mu +\frac{\rho _{r0}}{c^{2}}\left( \frac{%
a}{a_{0}}\right) ^{-4}\right) \right] ,  \nonumber
\end{equation}%
\[
\dot{\mu}+H(\mu -2\mu _{GX})=2H\frac{\rho _{r0}}{c^{2}}\left( \frac{a}{a_{0}}%
\right) ^{-4}, 
\]

\bigskip Solution for density in this model is

\[
\mu (a)=2\mu _{GX}+\frac{a_{0}}{a}(\mu _{0}-2\mu _{GX})+\frac{2}{3}\frac{%
\rho _{r0}}{c^{2}}\frac{a_{0}}{a}\left( 1-\left( \frac{a_{0}}{a}\right)
^{3}\right) , 
\]%
and its dependence on the scale factor is similar to the one shown at fig. (%
\ref{mod2}). This is because again with negligible contribution of radiation
the time derivative of the density is determined by the condition $\mu =2\mu
_{GX}$ which leads to the same density parameter $\Omega _{m}=2/3$.

The scale factor is the same as in the model 2 above. Radiation density parameter is
\[
\Omega_r=\frac{\pi^2}{3}\,\frac{\rho_{r0}}{\mu_{GX}a_0^2 H_0^2},
\]
and it coincides with (\ref{rhor}) due to definition of $\mu_{GX}$.

\textbf{Model 4, with radiation.} For this case cosmological equations read

\begin{equation}
H^{2}=\frac{8\pi G}{3}\left( \mu +\frac{1}{a}\sqrt{\frac{\pi }{8G\rho _{GX}}}%
\rho _{r0}\left( \frac{a}{a_{0}}\right) ^{-4}\right) +\frac{\beta }{a}, 
\nonumber
\end{equation}%
\[
\dot{\mu}+2H\mu =\frac{H}{a}\sqrt{\frac{\pi\rho_{GX}}{2G}}\left( 1+\frac{\rho _{r0}}{\rho _{GX}}\left( 
\frac{a}{a_{0}}\right) ^{-4}\right) , 
\]

\begin{figure}[htp]
	\centering
		\includegraphics{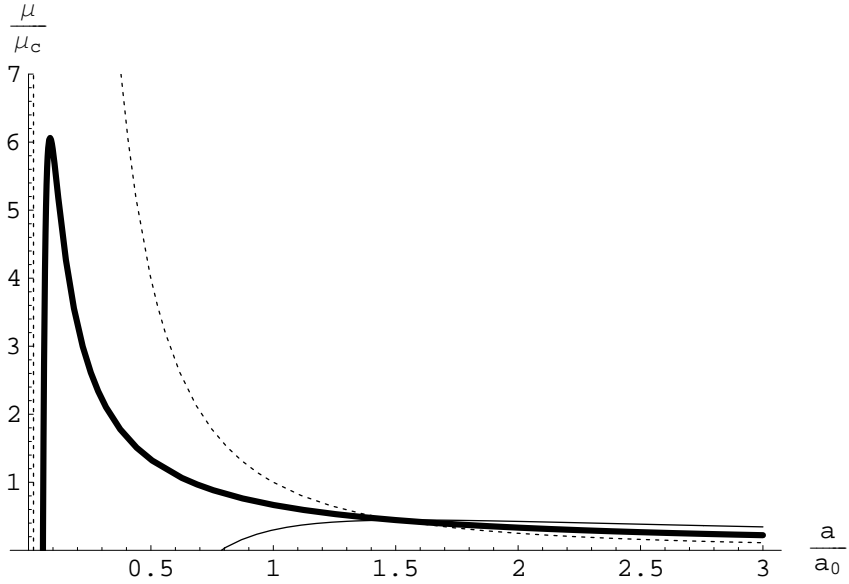}
\caption{Density of matter in model 4. $\Omega _{m}$ equals to 0.3 (firm curve), 0.667 (thick
curve) and 0.999 (dotted curve).}
\label{mod4}
\end{figure}

The solution for model 4 is

\[
\mu (a)=\mu _{0}\left( \frac{a_{0}}{a}\right) ^{2}+\sqrt{\frac{\pi \rho _{GX}%
}{2G}}\frac{1}{a}\left( 1-\frac{a_{0}}{a}\right) +\frac{\rho _{r0}a_{0}}{3}%
\sqrt{\frac{\pi }{2G\rho _{GX}}}\frac{1}{a^{2}}\left( 1-\left( \frac{a_{0}}{a}%
\right) ^{3}\right) , 
\]%
and it is represented at fig. (\ref{mod4}).

There is no special value of
density parameter in model 4. Solutions start with zero density and positive
scale factor, then the density increases rapidly in radiation dominated
stage, but entering into matter dominated stage it starts to decrease, so
for each solution there is a maximum in the density (see fig. (\ref{mod4})).

Here the density parameter of radiation is defined by
\[
\Omega_r=\frac{16\pi^{3/2}}{3H_0^2 a_0}\sqrt{\frac{G}{\rho_{GX}}}\rho_{r0},
\]
and it is again equivalent to (\ref{rhor}).

\section{Discussion and conclusions}

In this paper we explored various cosmological models with Gurzadyan-Xue
dark energy formula (\ref{rhoLambda}), assuming also variation of physical
constants such as the speed of light and the gravitational constant. It is
shown that unlike standard Friedmann models, various interesting
possibilities appear. In particular in the model 1 the continuity equation
is reduced to the energy transfer from dark energy component into the usual
matter. For some cases (models 1 and 4), the source term in the continuity
equation can be neglected for sufficiently large scale factors (or
sufficiently large present density of matter), and the discussion of the
dynamics in this asymptotic cases can be given in terms of the standard
cosmology.

Solutions for the density in terms of the scale factor for all models with
pressureless matter and dark energy contain separatrix which divides
solutions in two classes. The first class is analogous to the standard
Friedmann solutions with singularity in the beginning of expansion. The
second class, in contrast, does not contain singularity; solutions start
with vanishing density and non-zero scale factor.

We also considered generalization of GX models, including radiation,
assuming that dynamics of the latter is not influenced by the variation of
constants and by dark energy. Interestingly, for the model 1 we obtained the
behavior qualitatively similar to models without radiation. For the rest of
models separatrix in the sense discussed above is absent and all solutions
belong to the second class, namely non-Friedmannian solutions. The density
in these solutions may be larger in the past, but if so, it has a maximum
and all solutions start with vanishing density in radiation-dominated epoch.
However for models 2 and 3 with radiation the same value of density
parameter $\Omega _{m}=2/3$ that parametrizes separatrix in simple GX
models, corresponds to constant mass density of pressureless matter in
matter dominated epoch. For larger density parameter the mass density
decreases with expansion and for smaller density parameter it increases.

In fact, not all GX models with non-Friedmannian behavior ($\Omega_m<2/3$) pass age constraint and have deceleration parameter which fits the value followed from observations. At the same time model IV with $\Omega_m\geq 2/3$ pass both constraints and represent a good candidate for dark energy model \cite{VY06b} which in the limit of vanishing energy transfer from dark energy to the usual matter has $a^{-1}$ term in Friedmann equation. In the same way models I and III fit these constraints for density parameter slightly smaller than the separatrix value. This circumstance once more points out on a special role of the separatrix in GX models.

Notice another interesting feature of GX models with radiation. It is well known that Friedmannian models with zero curvature have arbitrary scale factor. The same holds also for simple GX models (model 2). However, this degeneracy breaks down in GX models with radiation, where the scale factor can always be computed in terms of present density parameters, and present values of (varying) physical constants.

We are thankful to the referees for helpful comments.

\end{document}